\documentclass[sn-mathphys-num]{sn-jnl}% Math and Physical Sciences Numbered Reference Style 
\usepackage{graphicx}%
\usepackage{multirow}%
\usepackage{amsmath,amssymb,amsfonts}%
\usepackage{amsthm}%
\usepackage{mathrsfs}%
\usepackage[title]{appendix}%
\usepackage{xcolor}%
\usepackage{textcomp}%
\usepackage{manyfoot}%
\usepackage{booktabs}%
\usepackage{algorithm}%
\usepackage{algorithmicx}%
\usepackage{algpseudocode}%
\usepackage{listings}%
\newtheorem{lemma}{Lemma}

\begin{document}

\title[SEM with heteroskedastic normal perturbations]{Spatial error models with heteroskedastic normal perturbations and joint modeling of mean and variance}

%%=============================================================%%
%% GivenName	-> \fnm{Joergen W.}
%% Particle	-> \spfx{van der} -> surname prefix
%% FamilyName	-> \sur{Ploeg}
%% Suffix	-> \sfx{IV}
%% \author*[1,2]{\fnm{Joergen W.} \spfx{van der} \sur{Ploeg} 
%%  \sfx{IV}}\email{iauthor@gmail.com}
%%=============================================================%%

\author[1]{\fnm{J.D.} \sur{Toloza}}\email{jdtolozad@unal.edu.co}

\author[1]{\fnm{O.O.} \sur{Melo}}\email{oomelo@unal.edu.co}

\author*[3]{\fnm{N.A.} \sur{Cruz}}\email{nelson-alirio.cruz@uib.es}

\affil[1]{Department of Statistics, Faculty of Sciences,  Universidad Nacional de Colombia, Bogota, Colombia}

\affil[2]{Universitat de les Illes Balears,  Departament de Matemàtiques i Informàtica, Phone: +34 637 54 6888, Palma de Mallorca, España }

\abstract{This work presents the spatial error model with heteroskedasticity, which allows the joint modeling of the parameters associated with both the mean and the variance, within a traditional approach to spatial econometrics. The estimation algorithm is based on the log-likelihood function and incorporates the use of GAMLSS models in an iterative form. Two theoretical results show the advantages of the model to the usual models of spatial econometrics and allow obtaining the bias of weighted least squares estimators.
The proposed methodology is tested through simulations, showing notable results in terms of the ability to recover all parameters and the consistency of its estimates. Finally, this model is applied to identify the factors associated with school desertion in Colombia.}
%%==================================%%
%% Sample for unstructured abstract %%
%%==================================%%
\keywords{Variance prediction, Generalized additive models, heteroscedasticity, Spatial regression,  Maximum likelihood estimation}

%%\pacs[JEL Classification]{D8, H51}

%%\pacs[MSC Classification]{35A01, 65L10, 65L12, 65L20, 65L70}

\maketitle
\section{Introduction}\label{sec1}

The continuous advance in information and communication technologies has caused a notable increase in the amount of data indexed in space, generating a growing interest in the scientific community to use appropriate tools for its analysis. This phenomenon has especially boosted the field of spatial econometrics, which has experienced considerable development in recent decades. This progress is reflected in creating and refining spatial regression algorithms available in specialized software such as R, STATA, and Python (\citealp{bivand2022r}).

In addition, spatial econometrics has made it possible to address different problems with economic data (\citealp{lopez2004regional}, \citealp{basile2014modeling}, and \citealp{martini2020regional}), social (\citealp{montero2018housing}, \citealp{delgado2021determinants}, and \citealp{canche2023spatial}) and environmental (\citealp{plant2018spatial}, and \citealp{ver2018spatial}). However, a key limitation of these models is the assumption of homoscedasticity, which is often violated in practice, leading to inefficient and biased estimates.

Although \cite{anselin1988spatial} formulated a general model that included heteroskedasticity in the errors and derived, from maximum likelihood, the first-order conditions and the information matrix, its computational implementation was not developed (\citealp{yokoi2010efficient}), as point that the packages currently used in R assume that homoscedasticity exists or apply robust errors in the estimation of the variance and covariance matrix (\citealp{bivand2005spdep}, \citealp{bivand2021review}, and \citealp{minguez2023package}).

In this sense, works such as those by \citet{lesage1997bayesian}, and \citet{cepeda2022spatial} stand out, who raise the possibility of incorporating variance modeling based on Bayesian methodologies. On the other hand, authors such as \citet{kelejian2010specification}, and \citet{arraiz2010spatial} address the problem of heteroscedasticity by using the generalized method of moments; this strategy is also implemented in R through the sphet library (\citealp{piras2010sphet}).

Despite these advances, the literature on spatial regression models with heteroscedasticity remains limited and fragmented. Most approaches adopted to date are Bayesian or based on hierarchical models, where the spatial component is modeled using a structured random effect (\citealp{lee2017data}, \citealp{lacombe2017hierarchical}, and \citealp{haining2020modelling}). This predominance leaves a void in the application of classical methods. Therefore, the present study aims to develop a spatial error model with heteroskedasticity based on the maximum likelihood method.

The proposed model is an extension of the traditional SEM that allows for heteroscedasticity in error variance, more accurately capturing the complexities of heterogeneous spatial data and improving the efficiency and accuracy of estimates. In addition, an estimation methodology based on GAMLSS (\citealp{rigby2005generalized}) models is presented, adapted from the work of \citet{toloza2020modelacion} for SAR models. Simulations and empirical applications are also carried out to validate the model.

The structure of the paper is as follows: in section 2, the theoretical formulation of the proposed model is presented and the estimation method is discussed, as well as some relevant results regarding the bias of the estimators. Section 3 details the simulation results that illustrate the effectiveness of the proposed model. Section four uses the proposed model on an empirical data set, demonstrating its applicability and advantages in real contexts. Finally, we conclude with a discussion of the findings and suggestions for future research in section five.

\section{SEM with non-homogeneous variance}

The SEM with heteroscedastic normal errors is a particular case of the general \citet{anselin1988spatial} model, where spatial dependence is considered through a spatial error model. Thus, the model can be expressed in matricial form as:

\begin{equation}
\mathbf{y}= \mathbf{X \boldsymbol{\beta}}+\boldsymbol{u}, \quad \boldsymbol{u}=\lambda\mathbf{Wu}+\boldsymbol{\epsilon}
\label{ECGSAR1}
\end{equation}
where $\boldsymbol{\epsilon} \sim N(\mathbf{0},\mathbf{\Omega})$, $\boldsymbol{\beta}$ is a vector of $(k+1) \times 1$ parameters of the $k$ explanatory variables, $|\lambda|<1$ is a autoregressive parameter and $\textbf{W}$ is a matrix of spatial weights, $\mathbf{\Omega}$ is the covariance matrix with diagonal elements $\mathbf{\Omega}_{ii}=h_i(\mathbf{Z}\boldsymbol{\alpha})$, $h_i>0$\footnote{the function $h()=\exp()$.}, $\mathbf{Z}$ is a matrix of explanatory variables for the variance and $\boldsymbol{\alpha}$ is a vector of $P \times 1$ parameters of $P-1$ explanatory variables. 

Equation \eqref{ECGSAR1} can be simplified taking into account that $\mathbf{B}=\mathbf{I}-\lambda\mathbf{W}$; in this sense, it can be reexpressed as:
\begin{equation}
\mathbf{y}= \mathbf{X \boldsymbol{\beta}}+\mathbf{B}^{-1}\boldsymbol{\epsilon}
\label{ECGSAR2}
\end{equation}

Next, the likelihood function is obtained, starting from the fact that the variance of the error is given by $\text{E}[\boldsymbol{\epsilon}\boldsymbol{\epsilon}^\top]=\mathbf{\Omega}$. There exits, a vector of homoscedastic errors $\mathbf{v}=\mathbf{\Omega}^{-1/2}\boldsymbol{\epsilon}$. In this way, from the equation \eqref{ECGSAR2}, the new error vector can be written as:
\begin{equation}
    \mathbf{v}=\mathbf{\Omega}^{-1/2}\mathbf{B}\left(\mathbf{y}-\mathbf{X}\boldsymbol{\beta}\right)
    \label{ECGSAR}
\end{equation}
Following \citet{anselin1988spatial}, although $\mathbf{v
}$ is a vector of independent errors with a standard normal distribution, these cannot be observed and the likelihood function will have to be based on $\mathbf{y}$. For this reason, the Jacobian is introduced, which allows the joint distribution of $\mathbf{y}$ to be derived from $\mathbf{v}$. Thus, using the equation \eqref{ECGSAR}, it is obtained that:
\begin{equation}
    J=\left|\frac{\partial\mathbf{v}}{\partial\mathbf{y}}\right|=|\mathbf{\Omega}^{-1/2}\mathbf{B}|=|\mathbf{\Omega}^{-1/2}||\mathbf{B}|=|\mathbf{\Omega}|^{-1/2}|\mathbf{B}|
    \label{EC4}
\end{equation}
Consequently, based on the standard normal distribution of the error term $\mathbf{v}$, and using the result of the equation \eqref{EC4}, the log-likelihood function for the observations vector $\mathbf{y}$ will be:

$$\ell=-\frac{n}{2}\ln(\pi)-\frac{1}{2}\ln|\mathbf{\Omega}|+\ln|\mathbf{B}|-\frac{1}{2}(\mathbf{y}-\mathbf{{X}\boldsymbol{\beta}})^\top\mathbf{B}^\top\mathbf{\Omega^{-1}}\mathbf{B}(\mathbf{y}-\mathbf{{X}\boldsymbol{\beta}})$$

To maximize this function, it must be performed using numeric methods (mainly the auto-regressive parameter $\lambda$). The method proposed corresponds to an adaptation of the algorithm proposed by \citet{anselin1988spatial} to estimate SEM models, which uses GAMLSS models to estimate the parameters associated with the mean and variance (\citealp{rigby2005generalized}). The methodology is based on the fact that by knowing $\lambda$ and multiplying the vector $\mathbf{y}$ and the matrix $\mathbf{X}$ by the matrix $\mathbf{B}=(\mathbf{I}-\lambda\mathbf{W})$, the model parameters can be obtained through a model for the mean and variance.
Thus, by deriving the log-likelihood function for the parameters of interest, the following equations are obtained:
\begin{align*}
\frac{\partial{\ell}}{\partial\boldsymbol{\beta}}&=\mathbf{v}^{\top}\mathbf{\Omega}^{-\frac{1}{2}}\mathbf{B}\mathbf{X}\\
\frac{\partial{\ell}}{\partial\lambda}&=-\operatorname{tr} (\mathbf{B}^{-1} \mathbf{W})+\mathbf{v}^{\top} \mathbf{\Omega}^{-1 / 2} \mathbf{W} (\mathbf{y}-\mathbf{X}\boldsymbol{\beta})\\
\frac{\partial{\ell}}{\partial\boldsymbol{\alpha}_p}&=-(1 / 2) \operatorname{tr} (\mathbf{\Omega}^{-1} \mathbf{H}_{{p}})+(1 / 2) \mathbf{v}^{\top} \mathbf{\Omega}^{-3 / 2} \mathbf{H}_{{p}}\mathbf{B}(\mathbf{y}-\mathbf{X} \boldsymbol{\beta})
\end{align*}
for $p=1,2,...,P$, where $\mathbf{H}_p$ is a diagonal matrix with elements $\frac{\partial\exp(\boldsymbol{\alpha}^{\top} \mathbf{z})}{\partial \alpha_p}$, where this last equation is equivalent to the score function obtained by \cite{aitkin1987modelling} when differentiating with respect to the vector of parameters associated with the variance.
The above system is nonlinear, mainly in the autoregressive parameter $\lambda$. Additionally, the estimation of the vector of $\boldsymbol{\alpha}$ is complicated since it depends on the $\boldsymbol{\beta}$, and these depend on $\lambda$. However, even though the system of equations is nonlinear, the vector of $\boldsymbol{\beta}$ has a closed solution and is given by:
\begin{equation}\label{beta}
\hat{\boldsymbol{\beta}}=\left(\mathbf{X}^{\top}\mathbf{B}^{\top}\mathbf{\Omega}^{-1}\mathbf{B}\mathbf{X}\right)^{-1}\mathbf{X}^{\top}\mathbf{B}^{\top}\mathbf{\Omega}^{-1}\mathbf{B}\mathbf{y}
\end{equation}

In this way, the following iterative algorithm is proposed based on the joint estimation of mean and variance modeled by GAMLSS:
\begin{enumerate}
    \item Estimate a GAMLSS for mean and variance between the dependent variable $\mathbf{y}$ and the explanatory variables ($\mathbf{{X}}$ and $\mathbf{{Z}}$). Estimate the variances $\hat{\sigma}_i^2$ to obtain $\hat{\mathbf{\Omega }}$.
   \item Construct the joint log-likelihood function $(\ell_c)$ and replace $\mathbf{\Omega}$ with the matrix $\hat{\mathbf{\Omega}}$, obtained in the previous step. The joint log-likelihood function is given by:
$$\ell_c=-\frac{n}{2}\ln(\pi)-\frac{1}{2}\ln|\hat{\mathbf{\Omega}}|+\ln|\mathbf{B}|- \frac{1}{2}(\mathbf{y}-\mathbf{{X}}\boldsymbol{\beta})^{\top}\mathbf{B}^{\top}\hat{\mathbf{\Omega}}^{-1}\mathbf{B}(\mathbf{y}-\mathbf{{X}}\boldsymbol{\beta})$$
where $$\hat{\boldsymbol{\beta}}=\left(\mathbf{X}^{\top}\mathbf{B}^{\top}\hat{\mathbf{\Omega}}^{-1}\mathbf{B}\mathbf{X}\right)^{-1}\mathbf{X}^{\top}\mathbf{B}^{\top}\hat{\mathbf{\Omega}}^{-1}\mathbf{B}\mathbf{y}$$

   \item Maximize the joint log-likelihood function and find $\hat{\lambda}$. 
   \item With the $\hat{\lambda}$ found in the previous step, the matrix $\mathbf{B}=(\mathbf{I}-\lambda\mathbf{W})$ is created and multiplied by $\mathbf{y}$ and $\mathbf{X}$ to obtain $\mathbf{By}$ and $\mathbf{BX}$.
    \item Taking into account that when estimating a GAMLSS using $\mathbf{By}$ as the dependent variable and $\mathbf{BX}$ as the matrix of explanatory variables, the estimators associated with the mean and variance are obtained since the matrix $\mathbf{ B}$ allows adjusting the estimators as if they were generalized least squares. In this way, the variances, $\hat{\sigma}_i^2$, to obtain a new version of $\hat{\mathbf{ \Omega}}$ are estimated.
    \item With the matrix $\hat{\mathbf{\Omega}}$, the joint log-likelihood function is constructed again, and maximize the joint log-likelihood function and find $\hat{\lambda}_f$.
 \item $\hat{\lambda}_f$ and $\hat{\lambda}$ are compared, and steps 5 to 7 are iterated until $\hat{\lambda}_f \approx \hat{\lambda}$
 \item  Given $\hat{\lambda}_f$, a GAMLSS model with the dependent variable given by $(\mathbf{I}-\hat{\lambda}_f\mathbf{W})\mathbf{y}$ and the matrix of explanatory variables given by $(\mathbf{I}-\hat{\lambda}_f\mathbf{W})\mathbf{X}$,  and the estimators $ \boldsymbol{\hat{\beta}}$ and $\boldsymbol{\hat{\alpha}}$ are obtained for the mean and variance, respectively.
\end{enumerate}
The proposed algorithm allows all parameters to be estimated jointly, which provides better estimates of $\lambda$. However, for large volumes of information, the calculation of $\ln|\mathbf{\Omega}|$ generates certain computational problems, since by directly carrying out the operation the software approaches it to infinity, preventing the log-likelihood function from being maximized. Therefore, using the Cholesky decomposition, the following equation is obtained:
$$\ln|\mathbf{\Omega}|=2\sum_{i=1}^n\ln\left[\text{diag}(\mathbf{L})_i\right]$$
where $\mathbf{L}$ is a triangular matrix and $\mathbf{L}^\top \mathbf{L}=\mathbf{\Omega}$.
The inference of the model is based on the results of \cite{anselin1988spatial}, from the Cramer Rao bound given by the inverse of the information matrix:

$$[\mathbf{\mathcal{I}}(\boldsymbol{\theta})]^{-1}=-E[\partial^2\ell/\partial\boldsymbol{\theta}\partial\boldsymbol{\theta}^{\top}]$$

In this way, the components of the information matrix are obtained from the second partial derivatives:
\begin{align*}
\mathcal{I}_{\boldsymbol{\beta} \boldsymbol{\beta}^{\top}}&=\mathbf{(BX)}^{\top}\mathbf{\Omega}^{-1}\mathbf{{BX}}\\
\mathcal{I}_{\boldsymbol{\beta} \lambda}&=\mathbf{0}\\
\mathcal{I}_{\boldsymbol{\beta} \boldsymbol{\alpha}^{\top}}&=\mathbf{0}\\
\mathcal{I}_{{\lambda}{\lambda}}&=\text{tr}(\mathbf{W}\mathbf{B}^{-1})^2+\textbf{tr}(\mathbf{\Omega}(\mathbf{W}\mathbf{B}^{-1})^\top\mathbf{\Omega}^{-1}(\mathbf{W}\mathbf{B}^{-1}))\\
\mathcal{I}_{{\lambda}{\boldsymbol{\alpha}_p}}&=\text{tr}(\mathbf{\Omega}^{-1}\mathbf{H}_p\mathbf{W}\mathbf{B}^{-1})\\
\mathcal{I}_{{\boldsymbol{\alpha}_p}{\boldsymbol{\alpha}_q}}&=\frac{1}{2}(\text{tr}(\mathbf{\Omega}^{-2}\mathbf{H}_p\mathbf{H}_q))
\end{align*}
where the last equation can be reexpressed as:
$$\mathcal{I}_{{\boldsymbol{\alpha}_p}{\boldsymbol{\alpha}_q}}=\frac{1}{2}(\mathbf{{Z}}^\top\mathbf{{Z}})$$
In this way, by inverting the information matrix and substituting the ML estimates parameters, the variance and covariance matrix of the estimators can be obtained, which allows for generating the confidence intervals and performing the relevant hypothesis tests, both for the parameters associated with the mean and those of the variance or standard deviation.
Furthermore, by exploring the properties of the proposed estimator, the following two results are obtained:
\begin{lemma}\label{lemma1}
    If the assumptions of the model \eqref{ECGSAR2} are true, then the estimator for $\pmb{\beta}$ given in Equation \eqref{beta} satisfies that:
    $$E(\hat{\pmb{\beta}}\vert \hat{\lambda}, \hat{\mathbf{\Omega}})=\pmb{\beta} + \mathcal{O}\left(\max\limits_{i=1, \ldots,n}\left\{-\frac{\hat{\lambda}}{\hat{\sigma}_i}+\frac{\lambda}{\sigma_i}\right\}\right) $$
\end{lemma}
\begin{proof}
    See Appendix A.
\end{proof}
The above result shows that the estimator is not unbiased, and its bias depends on $$\left(-\frac{\hat{\lambda}}{\hat{\sigma}_i}+\frac{\lambda}{\sigma_i} \right)$$ which measures the bias of both the parameter associated with the SEM part and the heteroscedasticity of the data. In case that the value of $\frac{\hat{\sigma}_i}{\sigma_i}$ is far from 1, the total bias of the estimator $\hat{{\beta}}$ will be large regardless of the bias of the estimator ${\lambda}$. Even if the estimator of $\lambda$ has very little bias, the bias given in Lemma \ref{lemma1} could be large when the variance estimators contain a large bias. In case that the estimate of $\sigma_i$ contains little bias, the bias of $\hat{\pmb{\beta}}$ will be affected by the quantity $ \frac{-\hat{\lambda}+\lambda}{\sigma_i}$, which could be large if $\sigma_i$ is small, and in case that $\sigma_i$ will be very large, the bias will decrease even with large biases of $\lambda$.
\begin{lemma}\label{lemma2}
    If the assumptions of the model \eqref{ECGSAR2} are true, and in addition, conditions 1)-5) defined in \cite{arraiz2010spatial} are met, the estimator for $\pmb{\beta}$ proposed in the equation \eqref {beta} satisfies that:
    $$Var(\hat{\pmb{\beta}}\vert \hat{\lambda}, \hat{\mathbf{\Omega}})\leq min\left[Var(\hat{\pmb{\beta}}_1\vert\hat{\lambda}), Var(\hat{\pmb{\beta}}_2\vert\hat{\lambda}), Var(\hat{\pmb{\beta}}_3\vert\hat{\lambda}, \hat{\rho}), Var(\hat{\pmb{\beta}}_4\vert\hat{\lambda}, \hat{\rho})\right]$$
    where $\hat{\pmb{\beta}}_1$ is the estimator of $\pmb{\beta}$ under a homoscedastic SEM \citep{anselin1988spatial}, $\hat{\pmb{\beta}}_2$ is the estimator of $\pmb{\beta}$ under a homoskedastic SAR model \citep{anselin1988spatial},  $\hat{\pmb{\beta}}_3$ is the estimator of $\pmb{\beta}$ under a homoscedastic SARAR model (\citealp{anselin1988spatial}), and $\hat{\pmb{\beta}}_4$ is the estimator of $\pmb{\beta}$ under a SARAR model robust to heteroscedasticity (\citealp{arraiz2010spatial}).
\end{lemma}
\begin{proof}
    See Appendix A.
\end{proof}
These two lemmas guarantee that the proposed estimator will give better results in inferential terms for a SEM with heteroscedasticity.

\section{Simulation Study}

A simulation study will be carried out to evaluate the proposed methodology. In each of the simulations, the parameters were estimated using the methodology previously built for each of these samples. In addition to assessing the performance against the original simulation parameters, the following models will be adjusted: i) Ho-SEM: maximum likelihood SEM with homoscedasticity (\citealp{anselin1988spatial}), ii) Proposed: SEM with heteroskedasticity (methodology proposed in this paper), iii) Ro-SEM: maximum likelihood SEM robust for heteroskedasticity, proposed in \citet{arraiz2010spatial}, iv) SARAR: SARAR model with robust estimation for heteroscedastic perturbations (\citealp{arraiz2010spatial}), and v) SAR: SAR model with homoscedastic perturbations.

Each $\pmb{y}=(y_1, \ldots, y_n)$ generated from a normal distribution on a regular grid was simulated, with $\pmb{\epsilon} \sim N(\pmb{0}, \pmb{\Omega})$, where $\pmb{\Omega} =\{ \omega_{ii}\}_{n\times n}$, and $\omega_{ii}=\sigma^2_i=\exp(\alpha_0+\alpha_1x_{2i}+\alpha_2x_{3i})$. Therefore, $\pmb{y}=\mathbf{X}\pmb{\beta}+\left(\mathbf{I}_n-\lambda\mathbf{W} \right)^{-1}\pmb{\epsilon}$, where $x_{1i}\sim N(0,1)$, $x_{2i}\sim N(2,1) $, $x_{3i}\sim U(0,1)$, $\mathbf{X}=(\mathbf{1}_n, \mathbf{x}_1, \mathbf{x}_2)$ and the $w_{ij}$ follow a tower-like first-order contiguity.
$i=1, \ldots, n$, $n=49, 81, 144, 400$, each value of $\lambda= -0.75, -0.5, -0.25, 0, 0.25, 0.5, 0.75$, $\beta_0=1$, $\beta_1=-1$, $\beta_2=0.5$, $\alpha_0=0,1$, $\alpha_1=-1,0$  and $\alpha_2=0,1$. 500 simulations of each combination of parameters were run. The estimators were carried out using the software \cite{Rmanual}, and codes are shown in the supplementary file \ref{sf1}. The simulation also allows us to observe the behavior of the models under homoscedasticity, which occurs when $\alpha_1=\alpha_2=0$.
 \begin{figure}[!ht]
\centering
\includegraphics[width=13cm]{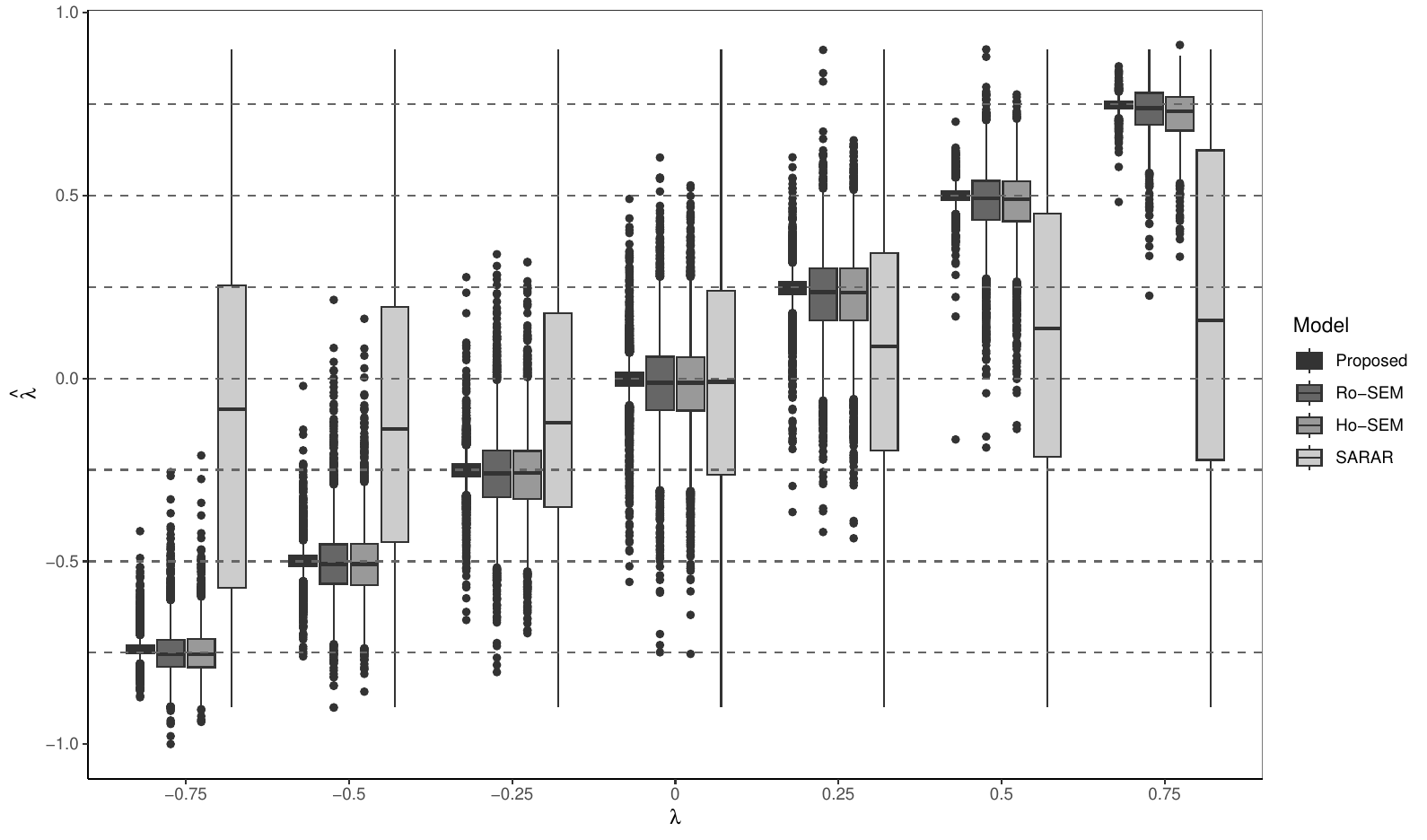}
\caption{$\hat{\lambda}$ for the different models adjusted for the simulation in a regular grid with $\beta_0=1$, $\beta_1=-1$, $\beta_2=0.5$, $\alpha_0=0,1$, $\alpha_1=-1,0$  and $\alpha_2=0,1$}
\label{rho1}
\end{figure}

Figure \ref{rho1} shows the estimators of $\lambda$, the spatial autoregressive parameter. It is noteworthy that for any true value of $\lambda$, the proposed model yields unbiased estimates, and a smaller variance than the other models. The robust autoregressive parameter estimation model (Ro-SEM) yields estimates with a slightly greater bias than the proposed model, although it slightly improves the homoscedastic SEM model, especially for large $\lambda$ values.
It is worth highlighting that the Ro-SEM has a similar variance to the Ho-SEM; therefore,  the model with robustness does not imply a large improvement in the estimation of $\lambda$ with respect to the homoscedastic model in terms of precision.

It can also be noted that the variance of the estimators decreases as the value of $\lambda$ moves away from 0. The SARAR model generates totally biased estimates of the $\lambda$ value, being a not very robust model both to heteroskedasticity, as it was explored in \cite{bivand2021review}, and to the misspecification of the autoregressive component.
Although the SARAR is at a disadvantage compared to the other models, it is evident that despite being planned to work robustly to heteroscedasticity, it fails to capture the spatial parameter in the error.
 \begin{figure}[!ht] 
\centering
\includegraphics[width=13cm]{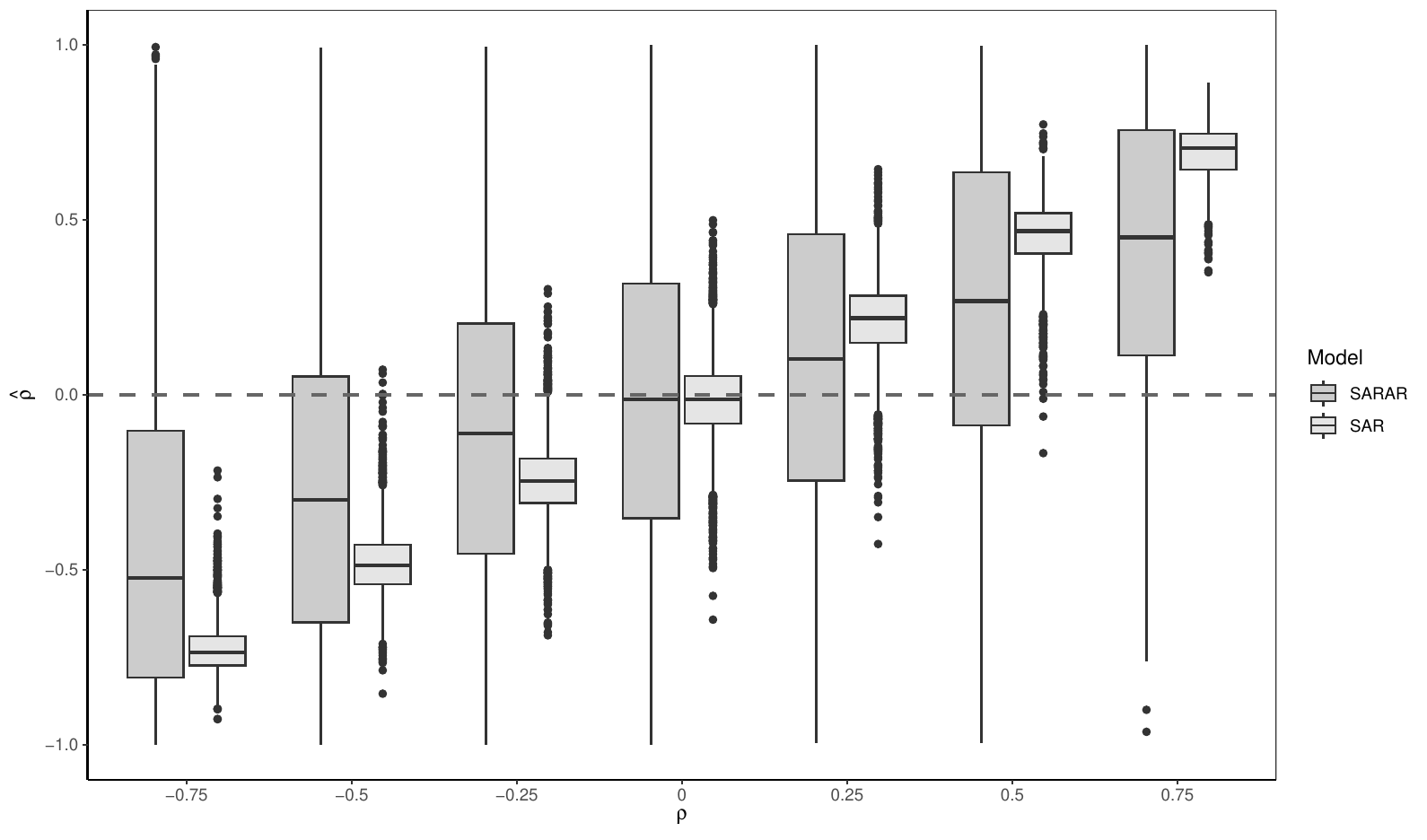}
\caption{$\hat{\rho}$ for the different models adjusted for the simulation in a regular grid with $\beta_0=1$, $\beta_1=-1$, $\beta_2=0.5$, $\alpha_0=0,1$, $\alpha_1=-1,0$  and $\alpha_2=0,1$ and $\lambda= -0.75, -0.5, -0.25, 0, 0.25, 0.5, 0.75.$}
\label{lambda1}
\end{figure}
The SARAR and SAR models estimate an autoregressive parameter ($\rho$) on the response, which was always $0$ in the simulation study, then the estimation of this value will be of interest. Figure \ref{lambda1} shows the estimates of the value of $\rho$ for the SAR and SARAR models. It is worth clarifying that the estimation method of the SARAR model from the library sphet (\citealp{piras2010sphet}) is based on the methodology of \cite{arraiz2010spatial}. It can be noted that the two models generate biased estimates of $\rho$ because they are affected by the SEM that generated the dataset. Also, the SAR model confuses the $\rho$ estimator with the value of $\lambda$, while the SARAR model presents a high variability of the $\rho$ estimator. Therefore, the results of the $\rho$ estimator show that SAR and SARAR are very sensitive to heteroscedasticity in the residuals when, in addition, there is a spatial error in the data-generating model.

 \begin{figure}[!ht] 
\centering
\includegraphics[width=0.8\textwidth]{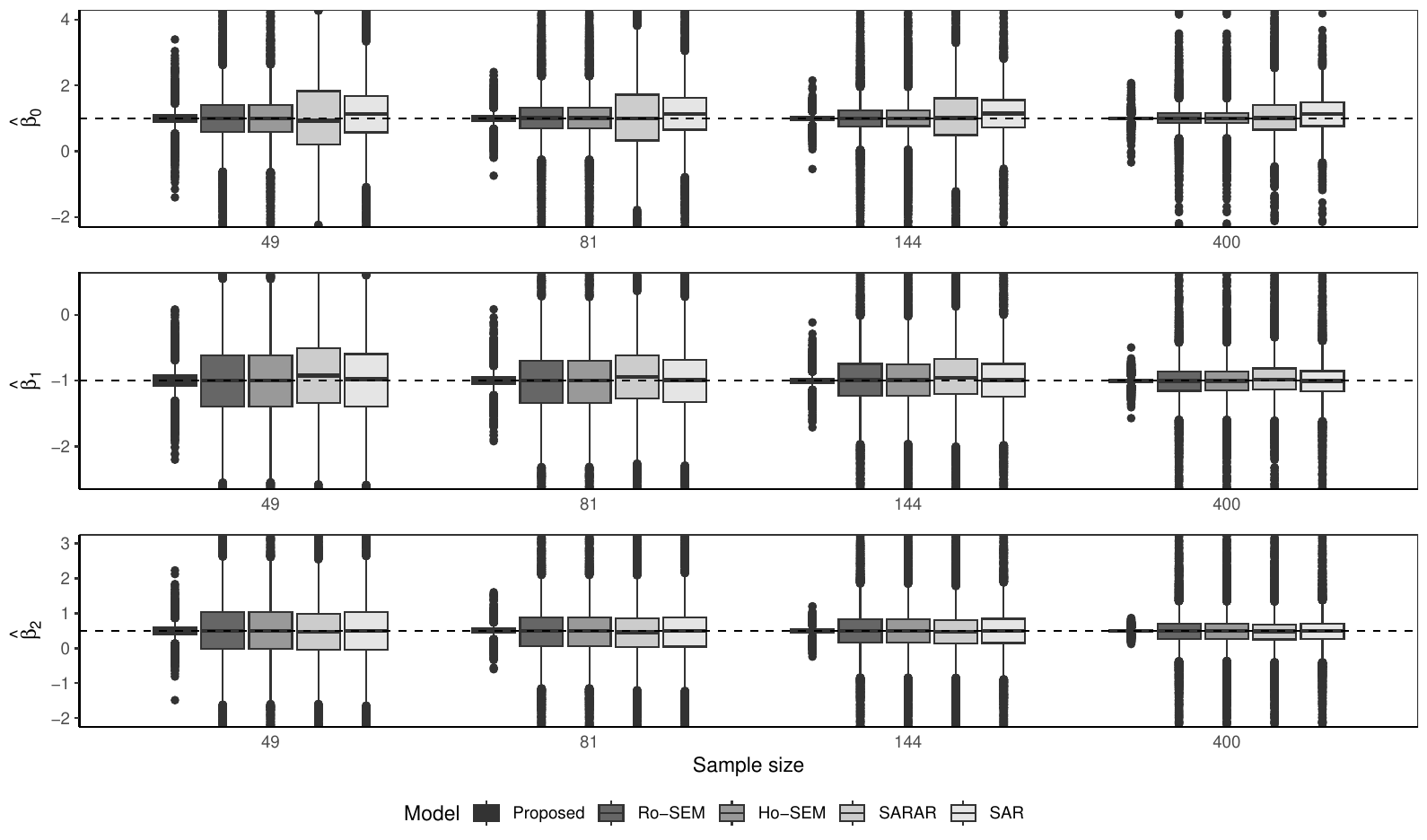}
\caption{$\hat{\beta}_0$, $\hat{\beta}_1$ and $\hat{\beta}_2$  for the different models adjusted for the simulation in a regular grid with $\beta_0=1$, $\beta_1=-1$, $\beta_2=0.5$, and at least one $\alpha_i\neq0$}
\label{b1}
\end{figure}
Figure \ref{b1} shows the estimators of the parameter $\beta_1=1$ in all simulation scenarios, when there is heteroskedasticity. Additionally, each subplot represents a different sample size. The proposed model has a very good performance compared to the other models, in terms of lower variance which shrinks rapidly as sample size grows. Although the SEM models present a higher variance than the proposed model, their estimators have the same variance and are unbiased. In addition, these models have a reduction in their variance as the sample size grows, but it is less pronounced than the proposed model.
This configuration raises the suspicion that the robust SEM does not improve the estimate of the parameter $\beta_1$ obtained by the homoskedastic SEM when there is heteroskedasticity in the error. It is worth highlighting that the SAR presents a bias that reduces as the sample size increases, a situation explained by the bias of the $\rho$ estimator, since the values of $\hat{\pmb{\beta}}$ are sensitive to this bias; a situation explained by \cite{santi2021reduced} and explored theoretically in \ref{lemma1}. However, the variance of the SAR model is lower, which is related to Figure \ref{lambda1}, which could show that when there is a heteroscedastic SEM as a data-generating process, a SAR is more robust than a SARAR.
 \begin{figure}[!ht] 
\centering
\includegraphics[width=13cm]{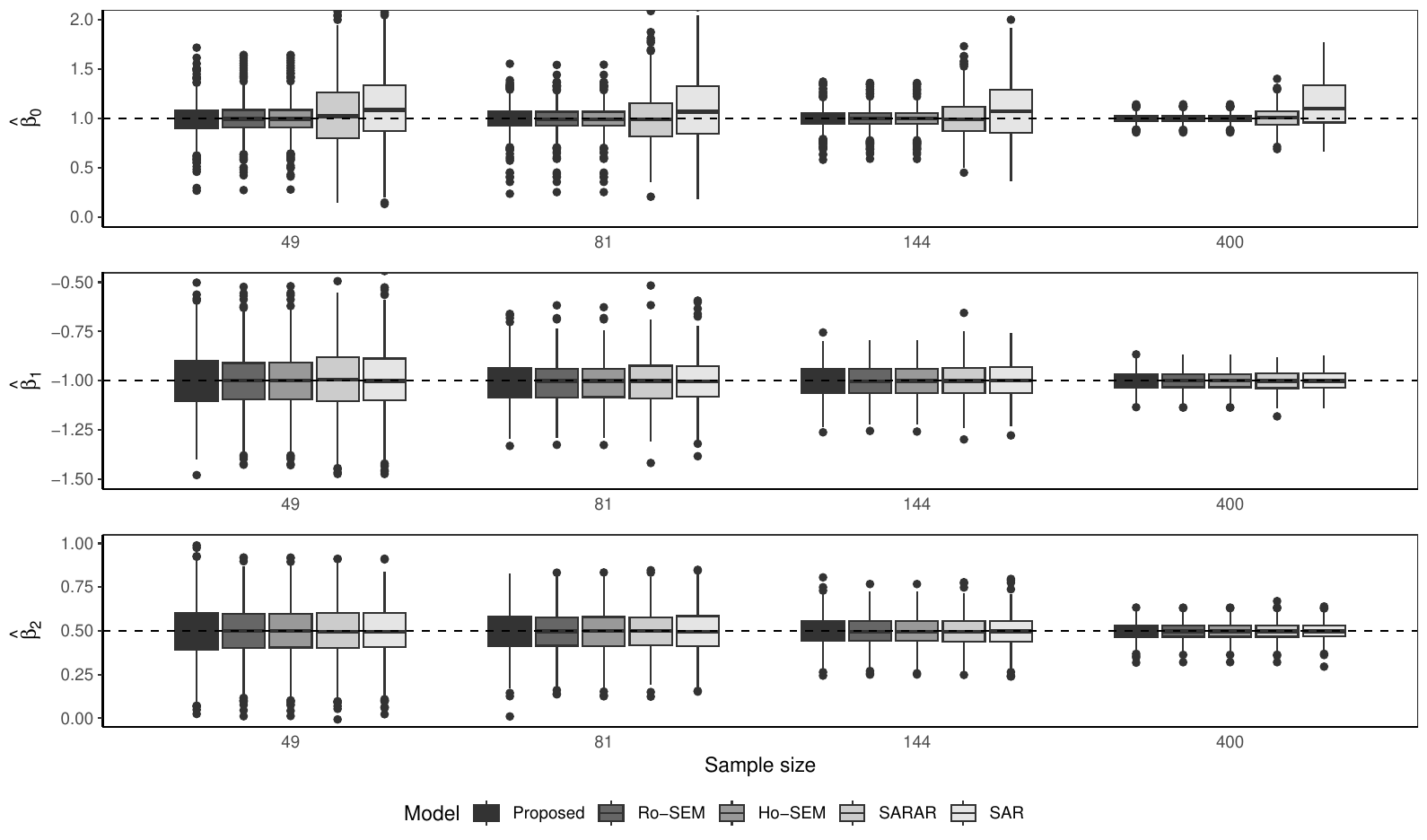}
\caption{$\hat{\beta}_0$, $\hat{\beta}_1$ and $\hat{\beta}_2$  for the different models adjusted for the simulation in a regular grid with $\beta_0=1$, $\beta_1=-1$, $\beta_2=0.5$,  $\alpha_1=0$  and $\alpha_2=0$}
\label{b2}
\end{figure}

Figure \ref{b2} shows the estimators of $\pmb{\beta}$ when there is homoscedasticity in the SEM, that is, $\alpha_1=0$ and $\alpha_2=0$. For $\beta_0$, the proposed model has a precision similar to the SEM-homoscedastic model and the robust SEM. The bias of the estimators is very low, except in the SAR model, which for the intercept overestimates the value of $\beta_0$. This could be explained by the bias committed when estimating $\rho$. Also, the SARAR model has a larger variance than the SEM models and the one proposed in this scenario, confirming the suspicion of its lack of robustness to the misspecification of the spatial component. For the estimates of the other two parameters, $\beta_1$ and $\beta_2$, the five models present very similar results, showing that under homoscedasticity, any model is adequate to estimate the effects of the variables on the response.

\begin{table}[!ht]
\footnotesize
\centering
\begin{tabular}{|c|c|ccc|ccc|ccc|ccc|}
  \hline
Model & n & \multicolumn{3}{c|}{Mean}& \multicolumn{3}{c|}{SD} & \multicolumn{3}{c|}{$P_{5\%}$}& \multicolumn{3}{c|}{$P_{95\%}$} \\ 
  & & $\beta_0$ &$\beta_1$ & $\beta_2$  &$\beta_0$ &$\beta_1$ & $\beta_2$&$\beta_0$ &$\beta_1$ & $\beta_2$&$\beta_0$ &$\beta_1$ & $\beta_2$ \\ 
  \hline
Proposed & 49 & 1.001 & -0.995 & 0.501 & 0.471 & 0.209 & 0.227 & 0.356 & -1.320 & 0.146 & 1.634 & -0.659 & 0.862 \\ 
  & 81 & 1.007 & -0.997 & 0.498 & 0.326 & 0.146 & 0.156 & 0.574 & -1.218 & 0.258 & 1.441 & -0.775 & 0.743 \\ 
  & 144 & 1.002 & -0.999 & 0.500 & 0.238 & 0.102 & 0.108 & 0.697 & -1.154 & 0.336 & 1.323 & -0.836 & 0.673 \\ 
   & 400 & 0.999 & -1.000 & 0.500 & 0.126 & 0.058 & 0.061 & 0.834 & -1.090 & 0.403 & 1.168 & -0.912 & 0.596 \\ 
   \hline
  Ro-SEM & 49 & 1.006 & -0.981 & 0.515 & 2.290 & 1.511 & 2.726 & -1.109 & -2.430 & -1.817 & 3.236 & 0.412 & 2.698 \\ 
   & 81 & 1.004 & -1.001 & 0.518 & 1.703 & 0.932 & 1.564 & -0.662 & -2.089 & -1.322 & 2.688 & 0.166 & 2.335 \\ 
 & 144 & 1.007 & -1.005 & 0.506 & 1.288 & 0.681 & 1.321 & -0.282 & -1.905 & -0.954 & 2.342 & -0.127 & 2.000 \\ 
 & 400 & 0.990 & -0.998 & 0.512 & 0.848 & 0.461 & 0.917 & 0.150 & -1.591 & -0.561 & 1.834 & -0.419 & 1.597 \\ 
 \hline
  Ho-SEM & 49 & 1.005 & -0.983 & 0.514 & 2.284 & 1.480 & 2.728 & -1.102 & -2.453 & -1.791 & 3.220 & 0.417 & 2.674 \\ 
   & 81 & 1.004 & -1.000 & 0.519 & 1.700 & 0.931 & 1.553 & -0.666 & -2.093 & -1.313 & 2.682 & 0.176 & 2.337 \\ 
  & 144 & 1.007 & -1.005 & 0.505 & 1.289 & 0.681 & 1.318 & -0.281 & -1.902 & -0.954 & 2.338 & -0.129 & 2.001 \\ 
 & 400 & 0.990 & -0.998 & 0.512 & 0.849 & 0.461 & 0.915 & 0.152 & -1.591 & -0.550 & 1.834 & -0.418 & 1.591 \\ 
 \hline
  SARAR & 49 & 0.981 & -0.924 & 0.488 & 1.954 & 1.650 & 2.641 & -1.098 & -2.374 & -1.668 & 3.596 & 0.506 & 2.610 \\ 
  & 81 & 1.028 & -0.949 & 0.504 & 1.495 & 0.927 & 1.523 & -0.709 & -2.008 & -1.238 & 3.242 & 0.253 & 2.268 \\ 
 & 144 & 1.010 & -0.956 & 0.485 & 1.254 & 0.697 & 1.274 & -0.432 & -1.863 & -0.973 & 2.882 & 0.000 & 1.951 \\ 
 & 400 & 0.997 & -0.964 & 0.502 & 0.875 & 0.476 & 0.914 & -0.209 & -1.545 & -0.557 & 2.396 & -0.298 & 1.592 \\ 
 \hline
  SAR & 49 & 1.033 & -0.967 & 0.512 & 1.358 & 1.655 & 2.679 & -0.732 & -2.448 & -1.839 & 2.817 & 0.442 & 2.768 \\ 
   & 81 & 1.053 & -0.998 & 0.519 & 1.035 & 0.930 & 1.624 & -0.444 & -2.137 & -1.357 & 2.449 & 0.187 & 2.393 \\ 
 & 144 & 1.025 & -1.006 & 0.506 & 0.865 & 0.696 & 1.386 & -0.155 & -1.931 & -1.020 & 2.148 & -0.106 & 2.053 \\ 
  & 400 & 1.017 & -0.996 & 0.513 & 0.670 & 0.473 & 0.954 & 0.060 & -1.606 & -0.600 & 1.964 & -0.402 & 1.668 \\ 
  \hline
\end{tabular}
\caption{Mean, standard deviation, 5\% percentile and 95\% percentile for estimated values of $\hat{\beta}_1$, $\hat{\beta}_2$ y $\hat{\beta}_3$ for the different models adjusted for the simulation in a regular grid with $\beta_0=1$, $\beta_1=-1$ and $\beta_2=0.5$, and overall $\alpha_j$ values  and different values of $n$}
\label{tablebeta1}
\end{table}

Table \ref{tablebeta1} shows some statistics of the estimation of each parameter $\pmb{\beta}$. In particular, the mean of the estimates, the standard deviation, the $5\%$ percentile, and the $95\%$ percentile are shown. This confirms that all models present unbiased results, but the model proposed in this work reduces the variance when there is the presence of heteroscedasticity in the response. Furthermore, in the presence of homoscedasticity, the estimates do not worsen with respect to other models.

If the columns of percentiles are also observed, it is noted that the tails of the distribution of the estimators under the proposed model are very light. While the tails of the other models are very heavy. For example, for $\beta_2=0.5$ and a sample size of $n=144$, the proposed model presents 90\% of all estimates in the interval $(0.336,0.673)$ while the best of the other four models presents 90\% in the interval $(-0.954, 2.001)$. This behavior demonstrates the goodness of the methodology proposed in this scenario in terms of inference.
 \begin{figure}[!ht] 
\centering
\includegraphics[width=13cm]{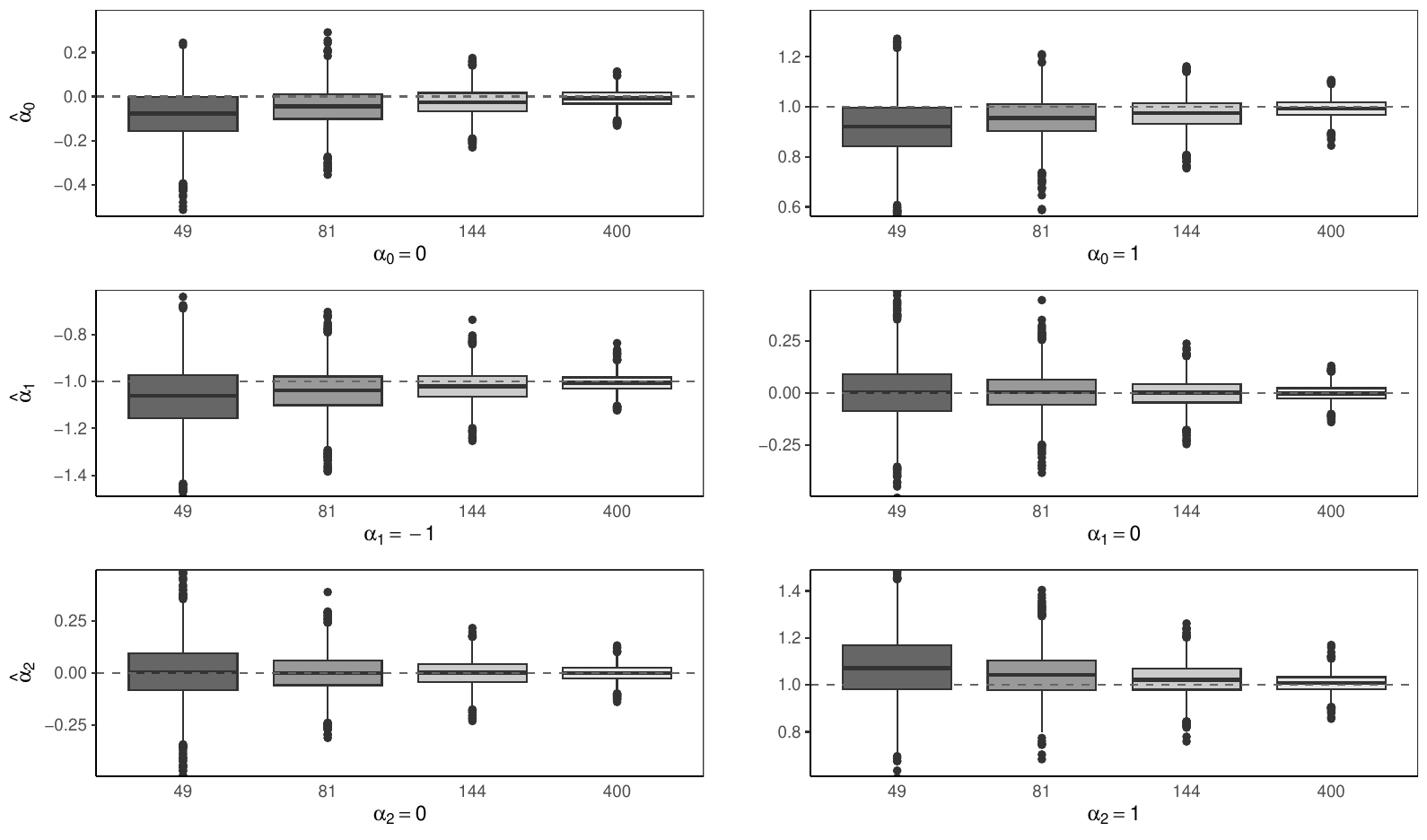}
\caption{$\hat{\alpha}_0$, $\hat{\alpha}_1$ and $\hat{\alpha}_2$ for the different models adjusted for the simulation in a regular grid with $\beta_0=1$, $\beta_1=-1$, $\beta_2=0.5$, $\alpha_0=0,1$, $\alpha_1=-1,0$  and $\alpha_2=0,1$. The x-axis represents the different sample sizes}
\label{a1}
\end{figure}

Finally, estimates of the effects of the variables on response variance are shown in Figure \ref{a1}. It is an additional advantage of the proposed model, which also allows the inclusion of variance modeling inherited from the GAMLSS methodology. The bias and variance of the estimates reduce as the sample size increases. In almost all cases, there is a bias with small sample sizes, but it is almost 0 when it approaches 400 observations in the grid.

\section{Application}

School desertion is a critical problem in many developing countries, including Colombia, and it has significant consequences for both individuals and society (\citealp{gomez2016desercion}). This phenomenon limits personal and professional development and perpetuates poverty and inequality (\citealp{roman2013factores}). Given the importance of addressing this problem, it is essential to understand the factors that contribute to school desertion to design and implement effective public policies. In this study, the proposed model is used, which allows for capturing both spatial dependence and heterogeneity in errors, providing a robust tool for the analysis and formulation of interventions aimed at reducing this variable.

\subsection{Data}

The database was built using the information on school desertion provided by the Ministry of National Education\footnote{Available at: https://www.datos.gov.co/}, which is disaggregated at the municipal level. On the other hand, for the explanatory variables, multidimensional synthetic indicators and variables consolidated by the National Planning Department (NPD) were used. These indicators and variables were obtained within the framework of the analyses carried out from the Modern Cities Index, developed by the NPD Cities System Observatory\footnote{Methodology and data available at: https://osc.dnp.gov. co/}. The variables used are defined below:
 \begin{figure}[!ht] 
\centering
\includegraphics[width=13cm]{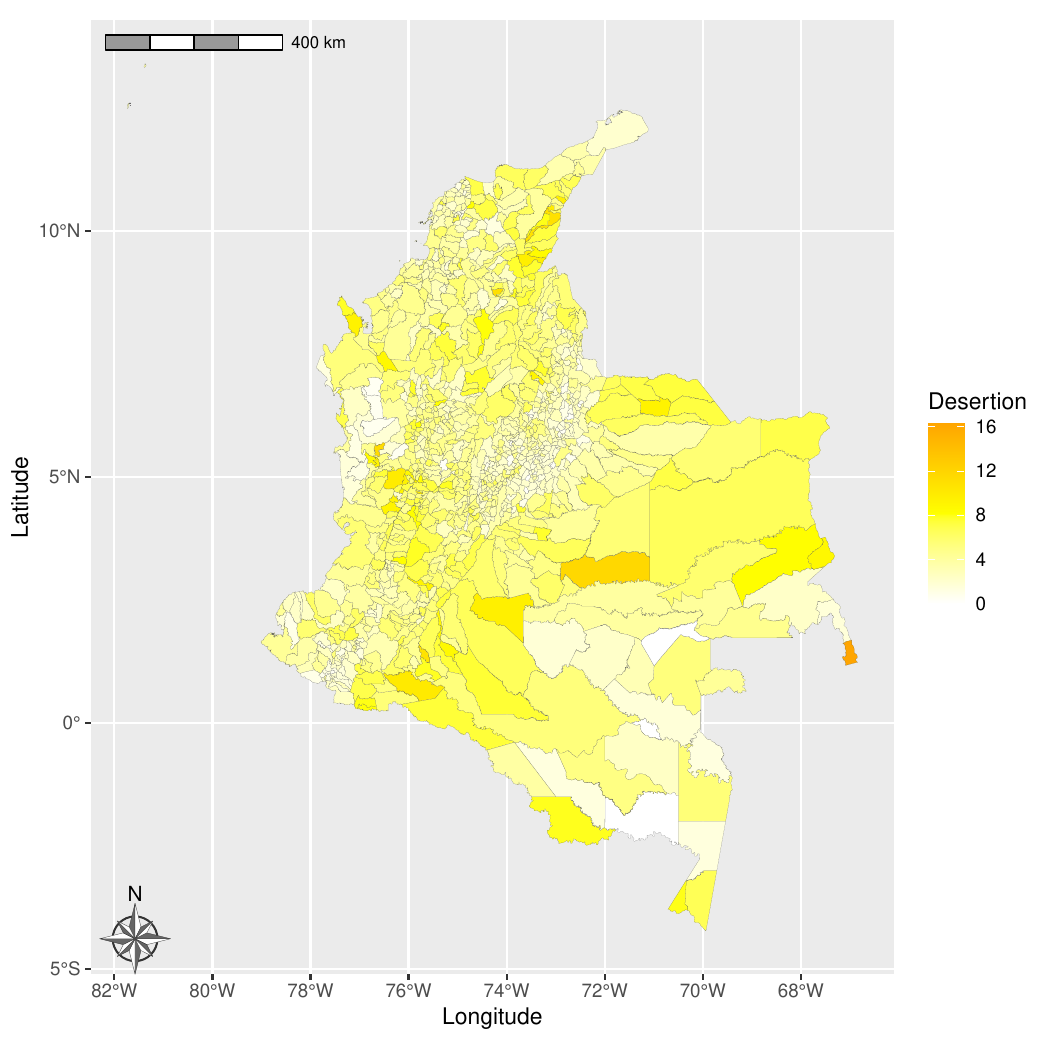}
\caption{Distribution of student desertion for 2022}
\label{map}
\end{figure}
\begin{itemize}
 \item Desertion $(y)$: Intra-annual dropout rate from the official sector. It identifies the proportion of enrolled students who, due to cultural, situational, or educational service provision factors, abandon their studies during the school year.
\item Victimization index $(x_1)$: The victimization risk index is an indicator that ranges from 0 to 1 and allows us to focus on those municipalities with the greatest number of cases of human rights violations.
\item Education index $(x_2)$: It is a synthetic indicator of the Equity and Social Inclusion dimension of the modern cities index that expresses performance in the Education domain, based on the use of the standardized variable “Standardized: Quality of education - Schools according to performance category” and “Standardized: education coverage rate”.
\item Homicide rate $(x_3)$: It is an indicator that refers to the number of deaths, due to causes related to homicides/murders, per 100000 inhabitants.
\item Hectares of forest deforested $(x_4)$: It is the inverse scaling in a range from 0 to 100 of the number of deforested hectares.
\end{itemize}
Figure \ref{map} shows the spatial behavior of school desertion for the year 2022, where a positive spatial correlation can be seen. This will be determined later from an exploratory and confirmatory analysis of spatial data.

\subsection{Results}

\begin{figure}[!ht] 
\centering
\includegraphics[width=13cm]{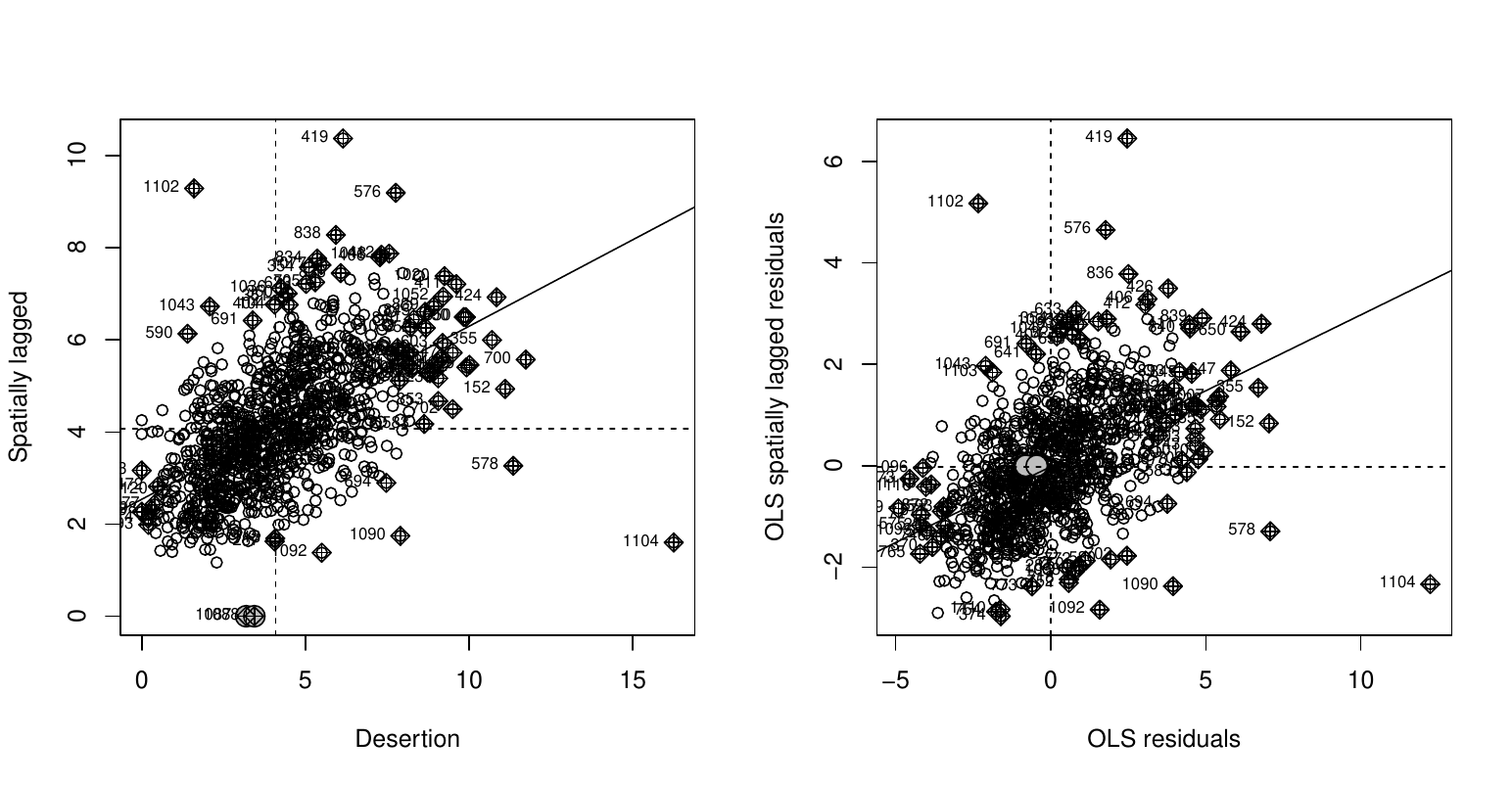}
\caption{Moran plot of desertion (left) and regression model residuals (right)}
\label{moran1}
\end{figure}

First, an exploratory analysis of spatial data is performed to determine the existence of spatial dependence. In this way, Figure \ref{moran1} presents the Moran graphs for the dropout variable and the residuals obtained from a traditional regression model with the following specification: $\hat{y}_i=\widehat{\beta}_0+\widehat{\beta}_1x_{1i}+\widehat{\beta}_2x_{2i}+\widehat{\beta}_3x_{3i}+\widehat{\beta}_4x_{4i}$. Graphically, a positive spatial correlation is observed in the data, which is confirmed with the Moran test, which yields significant results.

The model proposed for the estimation is:

$$\hat{y}_i=\widehat{\beta}_0+\widehat{\beta}_1x_{1i}+\widehat{\beta}_2x_{2i}+\widehat{\beta}_3x_{3i}+\widehat{\beta}_4x_{4i}$$

$$\widehat{\ln({\sigma}_i)}=\widehat{\alpha}_0+\widehat{\alpha}_1x_{1i}+\widehat{\alpha}_2x_{2i}+\widehat{\alpha}_3x_{3i}+\widehat{\alpha}_4x_{4i}$$

Table \ref{resulBogota} presents the results of the SEM (with homoscedasticity) and Proposed models, where it can be seen that there are slight differences in terms of the estimation of the parameter $\lambda$ and the $\boldsymbol{\beta}$. In general, the signs are the same under the two methodologies, although taking into account the results of the simulations, the proposed model's results should be more precise.

\begin{table}[!ht]
\centering
\footnotesize
\begin{tabular}{|ccc|}
  \hline
Parameter& Proposed & SEM \\ 
  \hline
 \multicolumn{3}{|c|}{Mean($\mu$)}\\
 \hline
$\hat{\beta}_0$ & 4.1003 ( 0.2021)  &  4.1384 (0.2429) \\ 
  $\hat{\beta}_1$ &1.7595 (0.4058) &  1.3725 (0.4839)  \\ 
  $\hat{\beta}_2$ &-0.0139 (0.0033)  & -0.0140 (0.0044)   \\ 
  $\hat{\beta}_3$&0.0066 ( 0.0013) & 0.0077 (0.0017)\\ 
 $\hat{\beta}_4$& 0.0002 (9.95$\times 10^{-5}$) &  0.0003 (0.0001) \\ 
 \hline
 \multicolumn{3}{|c|}{Variance  ($\sigma$)}\\
 \hline
 $\hat{\alpha}_0$ &-1.37670 (0.1563)&\\
  $\hat{\alpha}_1$ &-1.37670 (0.3061)&\\
  $\hat{\alpha}_2$ & -0.2248 (0.0029)& \\ 
  $\hat{\alpha}_2$ & 0.0232 (0.0013) &  \\ 
  $\hat{\alpha}_4$ & -0.3209 (0.0001) &  \\ 
  \hline
  $\lambda$ & 0.5949 (0.0315) & 0.5428 (0.0346) \\
  \hline
  MSE & 2.6225 & 2.6514   \\ 
   \hline
   \end{tabular}
   \caption{Results of the estimation (standard error in parenthesis) of the parametric effects over mean and standard
deviation for desertion for Proposed and SEM models}
   \label{resulBogota}
\end{table}
It was found that school desertion increase in municipalities with higher victimization rates, high homicide rates, and greater hectares of deforested\footnote{High values of this variable are related to municipalities with low institutional capacity.}. On the other hand, territorial entities with good results in education tend to have lower dropout rates. These findings are consistent with what was reported by \citet{gomez2016desercion}, who identified that factors such as poverty and violence are associated with high dropout rates. This aspect is especially relevant in Colombia, where there is a strong recruitment of young people by illegal groups (\citealp{diaz2019reclutamiento}). Furthermore, the variability has an inverse behavior with the education index and the homicide rate, while its relationship is directly concerning deforested hectares and the victimization index.
When performing a residual analysis of the SEM and proposed model, it is seen that both mitigate spatial dependence. However, the proposed model would have more precise estimates of the autoregressive parameter and the betas, as well as the ability to correctly estimate the behavior of the variance.
 \begin{figure}[!ht] 
\centering
\includegraphics[width=13cm]{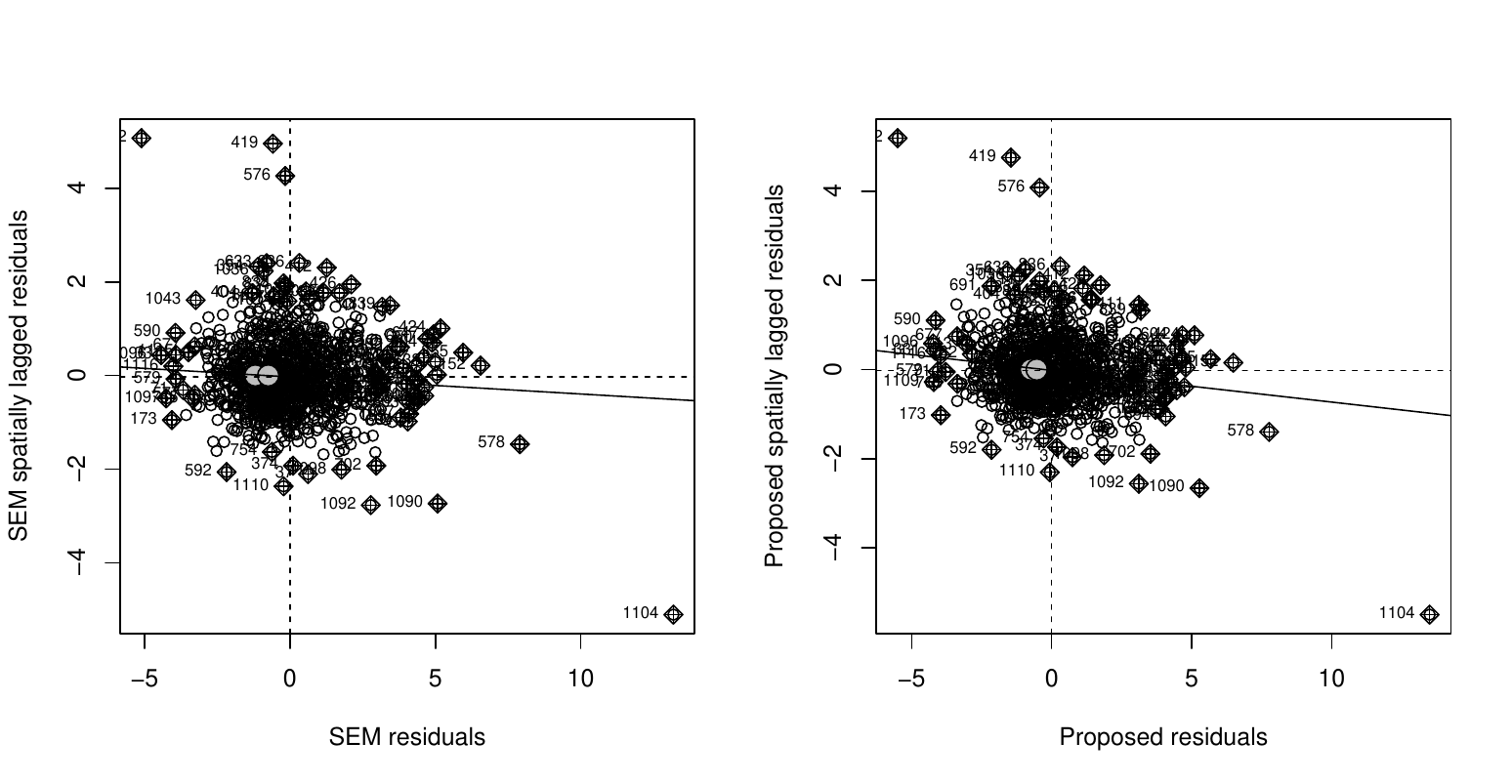}
\caption{Moran plot of SEM residuals (left) and proposed model residuals (right)}
\label{moran2}
\end{figure}

\section{Conclusions}
In this work, the heteroscedastic spatial error model was formulated, which allows the joint modeling of the parameters associated with the mean and variance, as well as the autoregressive spatial parameter of the error component. This methodology is based on the approaches of \cite{anselin1988spatial} and addresses the problem from a classic methodology of spatial econometrics based on maximum likelihood.

The results of the simulations carried out show that the proposed methodology allows us to more accurately obtain all the parameters involved in the data generating process, compared to other approaches, such as the traditional SEM or models that consider heteroskedasticity (\citealp{piras2010sphet}). In addition, the consistency of the estimates is highlighted, improving precision as the sample size increases.

Theoretically, an important result is derived: the variances of the $\hat{\boldsymbol{\beta}}$ will be less than or equal to those obtained by other methodologies, such as the SEM traditional, the robust SEM, the robust SARAR, and the traditional SARAR.

The model was applied to school dropout data in Colombia for the year 2022. It was found that school desertion decreased with the improvement of the educational quality of the municipality and increased in areas with high violence and deforestation. Likewise, variability reduces with a higher education index and a higher homicide rate, while it increases with more deforested hectares and a higher victimization rate.

For future work, the possibility of adjusting non-parametric or non-linear terms can be incorporated, as done by authors such as \citet{minguez2023package}, \citet{wood2017generalized}, and \citet{stasinopoulos2017flexible}. Furthermore, the algorithm can be extended to other spatial structures, such as spatial lag X, Durbin, or SARAR. Finally, although only four explanatory variables were used from the total NPD dataset, other specifications can be explored to understand school dropout behavior in Colombia better.

\backmatter

\bmhead{Supplementary information}

\begin{enumerate}
    \item Supplementary file 1\label{sf1}: R files with the simulations of this paper.
\end{enumerate}

\begin{appendices}

\section{Proof of Lemma \ref{lemma1}}\label{pl1}
\begin{proof}
  \begin{align*}
\hat{\boldsymbol{\beta}}\vert\hat{\lambda}, \hat{\mathbf{\Omega}}=&\left(\mathbf{X}^{\top}\hat{\mathbf{B}}^{\top}\hat{\mathbf{\Omega}}^{-1}\hat{\mathbf{B}}\mathbf{X}\right)^{-1}\mathbf{X}^{\top}\hat{\mathbf{B}}^{\top}\hat{\mathbf{\Omega}}^{-1}\hat{\mathbf{B}}\mathbf{y}\\
=&\bigg[ \mathbf{X}^\top \mathbf{B}^\top\mathbf{\Omega}^{-1}\mathbf{B}\mathbf{X} +  \mathbf{X}^\top \mathbf{B}^\top\mathbf{\Omega}^{-\frac{1}{2}}\mathbf{D}_{\hat{\lambda}, \hat{\mathbf{\Omega}}}\mathbf{W}\mathbf{X}+\\
&\mathbf{X}^\top \mathbf{W}^\top\mathbf{D}_{\hat{\lambda}, \hat{\mathbf{\Omega}}}\mathbf{\Omega}^{-\frac{1}{2}}\mathbf{B}\mathbf{X}+\mathbf{X}^\top \mathbf{W}^\top\mathbf{D}_{\hat{\lambda}, \hat{\mathbf{\Omega}}}\mathbf{D}_{\hat{\lambda}, \hat{\mathbf{\Omega}}}\mathbf{W}\mathbf{X}\bigg]^{-1}\\
&\left[\mathbf{X}^\top \mathbf{B}^\top\mathbf{\Omega}^{-\frac{1}{2}} +\mathbf{X}^\top \mathbf{W}^\top\mathbf{D}_{\hat{\lambda}, \hat{\mathbf{\Omega}}} \right]\left[\mathbf{\Omega}^{-\frac{1}{2}} \mathbf{B}+\mathbf{D}_{\hat{\lambda}, \hat{\mathbf{\Omega}}}\mathbf{W}\right]\left(\mathbf{X \boldsymbol{\beta}}+\mathbf{B}^{-1}\boldsymbol{\epsilon}\right)\\
=&\left[\mathbf{N}+\mathbf{M}\right]^{-1} \bigg[\mathbf{X}^\top \mathbf{B}^\top\mathbf{\Omega}^{-1}\mathbf{B}+\mathbf{X}^\top \mathbf{B}^\top\mathbf{\Omega}^{-\frac{1}{2}}\mathbf{D}_{\hat{\lambda}, \hat{\mathbf{\Omega}}}\mathbf{W}+\\
& \mathbf{X}^\top \mathbf{W}^\top\mathbf{D}_{\hat{\lambda}, \hat{\mathbf{\Omega}}}\mathbf{\Omega}^{-\frac{1}{2}} \mathbf{B}+\mathbf{X}^\top \mathbf{W}^\top\mathbf{D}^2_{\hat{\lambda}, \hat{\mathbf{\Omega}}}\mathbf{W}\bigg]\left(\mathbf{X \boldsymbol{\beta}}+\mathbf{B}^{-1}\boldsymbol{\epsilon}\right)
   \end{align*}
   where $\mathbf{N}=\mathbf{X}^\top \mathbf{B}^\top\mathbf{\Omega}^{-1}\mathbf{B}\mathbf{X}$ and, 
   \begin{align*}
       \mathbf{M}&=\mathbf{X}^\top \left[\mathbf{B}^\top\mathbf{\Omega}^{-\frac{1}{2}}\mathbf{D}_{\hat{\lambda}, \hat{\mathbf{\Omega}}}\mathbf{W}+\mathbf{W}^\top\mathbf{D}_{\hat{\lambda}, \hat{\mathbf{\Omega}}}\mathbf{\Omega}^{-\frac{1}{2}}\mathbf{B}+\mathbf{W}^\top\mathbf{D}^2_{\hat{\lambda}, \hat{\mathbf{\Omega}}}\mathbf{W}\right]\mathbf{X}\\
       \mathbf{D}_{\hat{\lambda}, \hat{\mathbf{\Omega}}}&=\text{diag}\left(-\frac{\hat{\lambda}}{\hat{\sigma}_i}+\frac{\lambda}{\sigma_i} \right)_{n\times n}
   \end{align*}
   If $\mathbf{N}$ is invertible, according to the inverse of a partitioned matrix in Theorem 3 of \citet[p.12]{magnus2019matrix} it is obtained that:
\begin{align*}
    \left(\mathbf{N} +\mathbf{M}\right)^{-1}&=\mathbf{N}^{-1}+\mathbf{N}^{-1}\mathbf{M}(\mathbf{I}+\mathbf{N}^{-1} \mathbf{M} )^{-1}\mathbf{N}^{-1} = \mathbf{N}^{-1}+\mathbf{Q}
\end{align*}
where $\mathbf{Q}=\mathbf{N}^{-1}\mathbf{M}(\mathbf{I}+\mathbf{N}^{-1} \mathbf{M} )^{-1}\mathbf{N}^{-1}$, and taking expectations with the assumption of $\boldsymbol{\epsilon} \sim N(\mathbf{0},\mathbf{\Omega})$ it is obtained that:
\begin{align}\nonumber
     E(\hat{\pmb{\beta}}\vert \hat{\lambda}, \hat{\rho}) & =\pmb{\beta}+\mathbf{QN}\pmb{\beta} +\left(\mathbf{N}^{-1} +\mathbf{Q}\right)\mathbf{X}^\top\bigg[\mathbf{B}^\top\mathbf{\Omega}^{-\frac{1}{2}}\mathbf{D}_{\hat{\lambda}, \hat{\mathbf{\Omega}}}\mathbf{W}+\nonumber\\
     &\mathbf{W}^\top\mathbf{D}_{\hat{\lambda}, \hat{\mathbf{\Omega}}}\mathbf{\Omega}^{-\frac{1}{2}}\mathbf{B}+
     \mathbf{W}^\top\mathbf{D}^2_{\hat{\lambda}, \hat{\mathbf{\Omega}}}\mathbf{W}\bigg]\mathbf{X}\boldsymbol{\beta} \nonumber\\
     &=\pmb{\beta} + \mathcal{O}\left(\max\limits_{i=1, \ldots,n}\left\{-\frac{\hat{\lambda}}{\hat{\sigma}_i}+\frac{\lambda}{\sigma_i}\right\}\right)
 \end{align}
   
\end{proof}
\section{Proof of Lemma \ref{lemma2}}\label{pl2}
\begin{proof}
Starting from the fact that the equation \eqref{ECGSAR2} is true, and furthermore, in the previous lemma  $\hat{\pmb{\beta}}-\pmb{\beta}=\left(\mathbf{X}^{\top}\hat{\mathbf{B}}^{\top}\mathbf{\Omega}^{-1}\hat{\mathbf{B}}\mathbf{X}\right)^{-1}\mathbf{X}^{\top}\hat{\mathbf{B}}^{\top}\mathbf{\Omega}^{-1}\hat{\mathbf{B}}\mathbf{B}^{-1}\pmb{\epsilon} +\mathcal{O}\left(\max\limits_{i=1, \ldots,n}\left\{-\frac{\hat{\lambda}}{\hat{\sigma}_i}+\frac{\lambda}{\sigma_i}\right\}\right)$, it is obtained that:
\begin{align*}
\hat{\pmb{\beta}}-\pmb{\beta}&=\left(\mathbf{X}^{\top}\hat{\mathbf{B}}^{\top}\mathbf{\Omega}^{-1}\hat{\mathbf{B}}\mathbf{X}\right)^{-1}\mathbf{X}^{\top}\hat{\mathbf{B}}^{\top}\mathbf{\Omega}^{-1}\hat{\mathbf{B}}\mathbf{B}^{-1}\pmb{\epsilon} +\mathcal{O}\left(\max\limits_{i=1, \ldots,n}\left\{-\frac{\hat{\lambda}}{\hat{\sigma}_i}+\frac{\lambda}{\sigma_i}\right\}\right)\\
&= \left(\mathbf{X}^{\top}\hat{\mathbf{B}}^{\top}\mathbf{\Omega}^{-1}\hat{\mathbf{B}}\mathbf{X}\right)^{-1}\mathbf{X}^{\top}\hat{\mathbf{B}}^{\top}\mathbf{\Omega}^{-1}\hat{\mathbf{B}}\mathbf{\hat{B}}^{-1}\pmb{\epsilon}  +\mathcal{O}\left(\max\limits_{i=1, \ldots,n}\left\{-\frac{\hat{\lambda}}{\hat{\sigma}_i}+\frac{\lambda}{\sigma_i}\right\}\right)\\
& +\left(\mathbf{X}^{\top}\hat{\mathbf{B}}^{\top}\mathbf{\Omega}^{-1}\hat{\mathbf{B}}\mathbf{X}\right)^{-1}\mathbf{X}^{\top}\hat{\mathbf{B}}^{\top}\mathbf{\Omega}^{-1}\hat{\mathbf{B}} \left(\mathbf{B}^{-1}-\mathbf{\hat{B}}^{-1}\right)\pmb{\epsilon} +\mathcal{O}\left(\max\limits_{i=1, \ldots,n}\left\{-\frac{\hat{\lambda}}{\hat{\sigma}_i}+\frac{\lambda}{\sigma_i}\right\}\right)\\
&= \left(\mathbf{X}^{\top}\hat{\mathbf{B}}^{\top}\mathbf{\Omega}^{-1}\hat{\mathbf{B}}\mathbf{X}\right)^{-1}\mathbf{X}^{\top}\hat{\mathbf{B}}^{\top}\mathbf{\Omega}^{-1}\pmb{\epsilon}+\mathcal{O}\left(\max\limits_{i=1, \ldots,n}\left\{-\frac{\hat{\lambda}}{\hat{\sigma}_i}+\frac{\lambda}{\sigma_i}\right\}\right) \\
& +\left(\mathbf{X}^{\top}\hat{\mathbf{B}}^{\top}\mathbf{\Omega}^{-1}\hat{\mathbf{B}}\mathbf{X}\right)^{-1}\mathbf{X}^{\top}\hat{\mathbf{B}}^{\top}\mathbf{\Omega}^{-1}\hat{\mathbf{B}} \left(\sum_{j=0}^\infty (\lambda-\hat{\lambda})^j\pmb{W}^j\right)\pmb{\epsilon}\\
\end{align*}
\begin{align*}
Var\left(\hat{\pmb{\beta}}\vert \hat{\lambda}\right)&=Var\Big[\left(\mathbf{\mathbf{\tilde{X}}}^{\top}\mathbf{\Omega}^{-1}\mathbf{\tilde{X}}\right)^{-1}\mathbf{\mathbf{\tilde{X}}}^{\top}\mathbf{\Omega}^{-1}\pmb{\epsilon}\\
& +\left(\mathbf{\mathbf{\tilde{X}}}^{\top}\mathbf{\Omega}^{-1}\mathbf{\tilde{X}}\right)^{-1}\mathbf{\mathbf{\mathbf{\tilde{X}}}^{\top}}\mathbf{\Omega}^{-1}\hat{\mathbf{B}} \left(\sum_{j=0}^\infty (\lambda-\hat{\lambda})^j\pmb{W}^j\right)\pmb{\epsilon}\Big]\\
&= \left(\mathbf{\mathbf{\tilde{X}}}^{\top}\mathbf{\Omega}^{-1}\mathbf{\tilde{X}}\right)^{-1}\mathbf{\mathbf{\tilde{X}}}^{\top}\mathbf{\Omega}^{-1}\mathbf{\Omega}\mathbf{\Omega}^{-1}\mathbf{\tilde{X}}\left(\mathbf{\mathbf{\tilde{X}}}^{\top}\mathbf{\Omega}^{-1}\mathbf{\tilde{X}}\right)^{-1}+\\
&\left(\mathbf{\mathbf{\tilde{X}}}^{\top}\mathbf{\Omega}^{-1}\mathbf{\tilde{X}}\right)^{-1}\mathbf{\mathbf{\mathbf{\tilde{X}}}^{\top}}\mathbf{\Omega}^{-1}\hat{\mathbf{B}} \left(\sum_{j=0}^\infty (\lambda-\hat{\lambda})^j\pmb{W}^j\right)\mathbf{\Omega}\left(\sum_{j=0}^\infty (\lambda-\hat{\lambda})^j\pmb{W}^j\right)^{\top} \\ &\mathbf{\hat{B}}\mathbf{\Omega}^{-1}\mathbf{\tilde{X}}\left(\mathbf{\mathbf{\tilde{X}}}^{\top}\mathbf{\Omega}^{-1}\mathbf{\tilde{X}}\right)^{-1}+\\
& 2\left(\mathbf{\mathbf{\tilde{X}}}^{\top}\mathbf{\Omega}^{-1}\mathbf{\tilde{X}}\right)^{-1}\mathbf{\mathbf{\tilde{X}}}^{\top}\mathbf{\Omega}^{-1}\mathbf{\Omega}\left(\sum_{j=0}^\infty (\lambda-\hat{\lambda})^j\pmb{W}^j\right)^{\top}\mathbf{\hat{B}}^{\top}\mathbf{\Omega}^{-1}\mathbf{\tilde{X}}\left(\mathbf{\mathbf{\tilde{X}}}^{\top}\mathbf{\Omega}^{-1}\mathbf{\tilde{X}}\right)^{-1}\\
&=\left(\mathbf{\mathbf{\tilde{X}}}^{\top}\mathbf{\Omega}^{-1}\mathbf{\tilde{X}}\right)^{-1}+\mathbf{R}_1(\lambda-\hat{\lambda})
\end{align*}
where $\mathbf{\tilde{X}}=\hat{\mathbf{B}}\mathbf{X}$, and 
\begin{align*}
    \mathbf{R}_1(\lambda-\hat{\lambda})&=\left(\mathbf{\mathbf{\tilde{X}}}^{\top}\mathbf{\Omega}^{-1}\mathbf{\tilde{X}}\right)^{-1}\mathbf{\mathbf{\mathbf{\tilde{X}}}^{\top}}\mathbf{\Omega}^{-1}\hat{\mathbf{B}} \left(\sum_{j=0}^\infty (\lambda-\hat{\lambda})^j\pmb{W}^j\right)\mathbf{\Omega}\left(\sum_{j=0}^\infty (\lambda-\hat{\lambda})^j\pmb{W}^j\right)^{\top} \\ &\mathbf{\hat{B}}\mathbf{\Omega}^{-1}\mathbf{\tilde{X}}\left(\mathbf{\mathbf{\tilde{X}}}^{\top}\mathbf{\Omega}^{-1}\mathbf{\tilde{X}}\right)^{-1}+\\
& 2\left(\mathbf{\mathbf{\tilde{X}}}^{\top}\mathbf{\Omega}^{-1}\mathbf{\tilde{X}}\right)^{-1}\mathbf{\mathbf{\tilde{X}}}^{\top}\left(\sum_{j=0}^\infty (\lambda-\hat{\lambda})^j\pmb{W}^j\right)^{\top}\mathbf{\hat{B}}^{\top}\mathbf{\Omega}^{-1}\mathbf{\tilde{X}}\left(\mathbf{\mathbf{\tilde{X}}}^{\top}\mathbf{\Omega}^{-1}\mathbf{\tilde{X}}\right)^{-1}\\
\end{align*}
Now, the estimator under the homoscedastic SEM is given by \citep{anselin1988spatial}:
\begin{align*}
    \hat{\pmb{\beta}}_1& = \left(\mathbf{X}^{\top}\hat{\mathbf{B}}^{\top}\hat{\mathbf{B}}\mathbf{X}\right)^{-1}\mathbf{X}^{\top}\hat{\mathbf{B}}^{\top}\hat{\mathbf{B}}\mathbf{y}=\pmb{\beta} + \left(\mathbf{X}^{\top}\hat{\mathbf{B}}^{\top}\hat{\mathbf{B}}\mathbf{X}\right)^{-1}\mathbf{X}^{\top}\hat{\mathbf{B}}^{\top}\pmb{\epsilon} \\
& +\left(\mathbf{X}^{\top}\hat{\mathbf{B}}^{\top}\hat{\mathbf{B}}\mathbf{X}\right)^{-1}\mathbf{X}^{\top}\hat{\mathbf{B}}^{\top}\hat{\mathbf{B}} \left(\sum_{j=0}^\infty (\lambda-\hat{\lambda})^j\pmb{W}^j\right)\pmb{\epsilon}\\
&=\pmb{\beta} + \left(\mathbf{\tilde{X}}^{\top}\mathbf{\tilde{X}}\right)^{-1}\mathbf{\tilde{X}}^{\top}\pmb{\epsilon}  +\left(\mathbf{\tilde{X}}^{\top}\mathbf{\tilde{X}}\right)^{-1}\mathbf{\tilde{X}}^{\top}\hat{\mathbf{B}} \left(\sum_{j=0}^\infty (\lambda-\hat{\lambda})^j\pmb{W}^j\right)\pmb{\epsilon}\\
\end{align*}
Therefore, it is obtained that:
\begin{align*}
    Var\left(\hat{\pmb{\beta}}_1|\hat{\lambda}\right) &= \left(\mathbf{\tilde{X}}^{\top}\mathbf{\tilde{X}}\right)^{-1}\mathbf{\tilde{X}}^{\top}\pmb{\Omega}\mathbf{\tilde{X}}\left(\mathbf{\tilde{X}}^{\top}\mathbf{\tilde{X}}\right)^{-1}+\\
    &\left(\mathbf{\mathbf{\tilde{X}}}^{\top}\mathbf{\tilde{X}}\right)^{-1}\mathbf{\mathbf{\mathbf{\tilde{X}}}^{\top}}\hat{\mathbf{B}} \left(\sum_{j=0}^\infty (\lambda-\hat{\lambda})^j\pmb{W}^j\right)\mathbf{\Omega}\left(\sum_{j=0}^\infty (\lambda-\hat{\lambda})^j\pmb{W}^j\right)^{\top} \mathbf{\hat{B}}^{\top}\mathbf{\tilde{X}}\left(\mathbf{\mathbf{\tilde{X}}}^{\top}\mathbf{\tilde{X}}\right)^{-1}+\\
& 2\left(\mathbf{\mathbf{\tilde{X}}}^{\top}\mathbf{\tilde{X}}\right)^{-1}\mathbf{\mathbf{\tilde{X}}}^{\top}\mathbf{\Omega}\left(\sum_{j=0}^\infty (\lambda-\hat{\lambda})^j\pmb{W}^j\right)^{\top}\mathbf{\hat{B}}^{\top}\mathbf{\tilde{X}}\left(\mathbf{\mathbf{\tilde{X}}}^{\top}\mathbf{\tilde{X}}\right)^{-1}\\
&= \left(\mathbf{\mathbf{\tilde{X}}}^{\top}\mathbf{\Omega}^{-1}\mathbf{\tilde{X}}\right)^{-1}+\mathbf{R}_1(\lambda-\hat{\lambda}) +  \mathbf{U}^{\top}\pmb{\Omega}\mathbf{U} + \\
&\mathbf{U}^{\top}\hat{\mathbf{B}} \left(\sum_{j=0}^\infty (\lambda-\hat{\lambda})^j\pmb{W}^j\right)\mathbf{\Omega}\left(\sum_{j=0}^\infty (\lambda-\hat{\lambda})^j\pmb{W}^j\right)^{\top} \mathbf{\hat{B}}^{\top}\mathbf{U} \\
&= Var\left(\hat{\pmb{\beta}}\vert \hat{\lambda}\right) + \mathbf{U}^{\top}\pmb{\Omega}\mathbf{U} + \\
&\mathbf{U}^{\top}\hat{\mathbf{B}} \left(\sum_{j=0}^\infty (\lambda-\hat{\lambda})^j\pmb{W}^j\right)\mathbf{\Omega}\left(\sum_{j=0}^\infty (\lambda-\hat{\lambda})^j\pmb{W}^j\right)^{\top} \mathbf{\hat{B}}^{\top}\mathbf{U} \\
& \geq Var\left(\hat{\pmb{\beta}}\vert \hat{\lambda}\right)
\end{align*}
where
\begin{align*}
\mathbf{U}&=\mathbf{\tilde{X}}\left(\mathbf{\mathbf{\tilde{X}}}^{\top}\mathbf{\tilde{X}}\right)^{-1}-\mathbf{\Omega}^{-1}\mathbf{\tilde{X}}\left(\mathbf{\mathbf{\tilde{X}}}^{\top}\mathbf{\Omega}^{-1}\mathbf{\tilde{X}}\right)^{-1}
\end{align*}
Assuming that the value $\lambda-\hat{\lambda}$ is the same for the proposed model as for the homoscedastic SEM; however, in \citet{arbia2006spatial}, and \citet{arraiz2010spatial}, it is discussed that in the homoscedastic SEM, $\lambda-\hat{\lambda}$ is larger when homoscedasticity is assumed.
Now, under a homoscedastic SAR following the result of appendix A.1 of \citet{santi2021reduced}, and adapting with $\rho=0$, it is obtained that:
\begin{align*}
    \hat{\pmb{\beta}}_2 &=  \pmb{\beta} + \left(\mathbf{X}^{\top}\mathbf{X}\right)^{-1}\mathbf{X}^{\top}\mathbf{B}^{-1}\pmb{\epsilon}+\hat{\rho}\left(\mathbf{X}^{\top}\mathbf{X}\right)^{-1}\mathbf{X}^{\top}\mathbf{W}\mathbf{B}^{-1}\pmb{\epsilon}
\end{align*}
Therefore, it is obtained that: 
\begin{align*}
    Var\left(\hat{\pmb{\beta}}_2\vert \hat{\rho}\right)&=\left(\mathbf{X}^{\top}\mathbf{X}\right)^{-1}\mathbf{X}^{\top}\mathbf{B}^{-1}\mathbf{\Omega} \mathbf{B^{\top}}^{-1}\mathbf{X}\left(\mathbf{X}^{\top}\mathbf{X}\right)^{-1}+\\
&\hat{\rho}^2\left(\mathbf{X}^{\top}\mathbf{X}\right)^{-1}\mathbf{X}^{\top}\mathbf{W}\mathbf{B}^{-1}\mathbf{\Omega} \mathbf{B^{\top}}^{-1}\mathbf{W}\mathbf{X}\left(\mathbf{X}^{\top}\mathbf{X}\right)^{-1}+\\
&2\hat{\rho}\left(\mathbf{X}^{\top}\mathbf{X}\right)^{-1}\mathbf{X}^{\top}\mathbf{B}^{-1}\mathbf{B^{\top}}^{-1}\mathbf{W}\mathbf{X}\left(\mathbf{X}^{\top}\mathbf{X}\right)^{-1}\\
& =\left(\mathbf{\mathbf{\tilde{X}}}^{\top}\mathbf{\Omega}^{-1}\mathbf{\tilde{X}}\right)^{-1}+\mathbf{R}_1(\lambda-\hat{\lambda}) +  \mathbf{U_2}^{\top}\mathbf{B}^{-1}\mathbf{\Omega} \mathbf{B^{\top}}^{-1}\mathbf{U_2} + \\
&\hat{\rho}^2\mathbf{U_2}^{\top}\mathbf{W}\mathbf{B}^{-1}\mathbf{\Omega} \mathbf{B^{\top}}^{-1}\mathbf{W}\mathbf{U_2} \\
&= Var\left(\hat{\pmb{\beta}}\vert \hat{\lambda}\right) \mathbf{U_2}^{\top}\mathbf{B}^{-1}\mathbf{\Omega} \mathbf{B^{\top}}^{-1}\mathbf{U_2} + \hat{\rho}^2\mathbf{U_2}^{\top}\mathbf{W}\mathbf{B}^{-1}\mathbf{\Omega} \mathbf{B^{\top}}^{-1}\mathbf{W}\mathbf{U_2} \\
& \geq Var\left(\hat{\pmb{\beta}}\vert \hat{\lambda}\right)
\end{align*}
where $\mathbf{U_2}=\mathbf{{X}}\left(\mathbf{\mathbf{{X}}}^{\top}\mathbf{{X}}\right)^{-1}-\mathbf{\Omega}^{-1}\mathbf{{X}}\left(\mathbf{\mathbf{{X}}}^{\top}\mathbf{\Omega}^{-1}\mathbf{{X}}\right)^{-1}$. The estimator proposed by \citet{arraiz2010spatial} for $\pmb{\beta}$ with a robust SARAR is more efficient than the one proposed by \citet{anselin1988spatial} for a homoscedastic SARAR, it is already clear that $Var(\hat{\pmb{\beta}}_3\vert\hat{\lambda}, \hat{\rho})\geq Var(\hat{\pmb{\beta}}_4)\vert\hat{\lambda}, \hat{\rho})$. Therefore, $\hat{\pmb{\beta}}_4$ will be analyzed, which is given by
\begin{align*}
    \hat{\pmb{\beta}}_4 &= \left[\mathbf{X_W}^{\top}(\mathbf{I}-\hat{\rho}\mathbf{W})^{\top}\mathbf{P}_H(\mathbf{I}-\hat{\rho}\mathbf{W})\mathbf{X_W})\right]^{-1}\mathbf{X_W})^{\top}(\mathbf{I}-\hat{\rho}\mathbf{W})^{\top}\mathbf{P}_H(\mathbf{I}-\hat{\rho}\mathbf{W})\mathbf{y}
\end{align*}
where $\mathbf{X_W}=[\mathbf{X}, \mathbf{W}]$, that is, a matrix containing the rows of $\mathbf{X}$ and the rows of $\mathbf{W}$, $\mathbf{P}_H=\mathbf{H(H^{\top}H)^{-1}H}$, with $\mathbf{H}=[\mathbf{X}, \mathbf {WX}]$, if $\mathcal{C}(\mathbf{A})$ is the column space of the matrix $\mathbf{A}$, then it is obtained that:
\begin{itemize}
 \item[1)] $\mathcal{C}(\mathbf{X}) \subset \mathcal{C}([\mathbf{X}, \mathbf{W}])$ Corollary 4.2.2 of \citet{harville1998matrix}
 \item[2)] $\mathcal{C}(\mathbf{WX}) \subset \mathcal{C}(\mathbf{X})$ Corollary 4.2.3 of \citet{harville1998matrix}
 \item[3)] $\mathcal{C}(\mathbf{W}) \subset \mathcal{C}([\mathbf{X}, \mathbf{W}])$ Corollary 4.2.3 of \citet{harville1998matrix}
 \item[4)] $\mathcal{C}(\mathbf{PH}) \subset \mathcal{C}(\mathbf{W})$ by 2) and Theorem 12.3.4 of \citet{harville1998matrix}
\end{itemize}
 By theorem 12.3.5 of \citet{harville1998matrix} and numerals 1) to 4), it follows that $\hat{\pmb{\beta}}_4$ will be less efficient than $\hat{\pmb{\beta}} $ while the column space of $\mathcal{C}([\mathbf{X}, \mathbf{W}])\subset\mathcal{C}([\mathbf{X}, \mathbf{W}, \mathbf {Z}])$. Therefore, if the variance specification for the SEM defined in the equation \eqref{ECGSAR2} is correct, the estimator proposed in the equation \eqref{beta} will always satisfy that $Var(\hat{\pmb{\beta}}_4\vert\hat{\lambda}, \hat{\rho})\geq Var(\hat{\pmb{\beta}})\vert\hat{\lambda})$
    
\end{proof}
\end{appendices}

%%===========================================================================================%%
%% If you are submitting to one of the Nature Portfolio journals, using the eJP submission   %%
%% system, please include the references within the manuscript file itself. You may do this  %%
%% by copying the reference list from your .bbl file, paste it into the main manuscript .tex %%
%% file, and delete the associated \verb+\bibliography+ commands.                            %%
%%===========================================================================================%%

\end{document}